\documentclass[aps, twocolumn, pre, superscriptaddress, floatfix]{revtex4-1}
\usepackage{graphicx, natbib,titlesec,amssymb, amsmath, paralist, hyperref}
\pdfoutput=1 

\begin{document}
\title{Fixation Probabilities in Weakly Compressible Fluid Flows}

\author{Abigail Plummer}
\email{plummer@g.harvard.edu}
\affiliation{Department of Physics, Harvard University, Cambridge, MA 02138}
\author{Roberto Benzi} 
\affiliation{Department of Physics and Istituto Nazionale di Fisica Nucleare, University of Rome Tor Vergata, 00133 Rome, Italy}
\author{David R. Nelson}
\affiliation{Department of Physics, Harvard University, Cambridge, MA 02138}
\author{Federico Toschi}
\affiliation{Department of Applied Physics and Department of Mathematics and Computer Science, Eindhoven University of Technology, 5600 MB Eindhoven, The Netherlands}
\affiliation{Istituto per le Applicazioni del Calcolo, Consiglio Nazionale delle Ricerche, 00185 Rome, Italy}
\date{January 8, 2019}

\begin{abstract}
Competition between biological species in marine environments is affected by the motion of the surrounding fluid. An effective 2D compressibility can arise, for example, from the convergence and divergence of water masses at the depth at which passively traveling photosynthetic organisms are restricted to live. In this report, we seek to quantitatively study genetics under flow. To this end, we couple an off-lattice agent-based simulation of two populations in 1D to a weakly compressible velocity field--first a sine wave and then a shell model of turbulence. We find for both cases that even in a regime where the overall population structure is approximately unaltered, the flow can significantly diminish the effect of a selective advantage on fixation probabilities. We understand this effect in terms of the enhanced survival of organisms born at sources in the flow and the influence of Fisher genetic waves.
\end{abstract}

\maketitle

Oceanic flows can affect competition between marine species in important ways, particularly at the submeso- and mesoscale where the characteristic timescales of fluid motion are comparable to the generation time (inverse growth rate) of phytoplankton \cite{dovidio, levy, prairie}. Recent observational and computational work on marine fronts, such as boundary currents and upwelling regions, has noted the likely importance of strong vertical velocities and turbulent eddies to the high productivity and genetic diversity of these regions \cite{levy2012, clayton2014, clayton2013, demonte}.

However, few quantitative connections have been made between the observational and numerical data on fronts and the literature of population genetics, which often relies on simplifying assumptions such as a constant carrying capacity, discrete subpopulations, no vertical mixing, and/or fixed migration patterns \cite{wares, chust2013, giovannoni2005, krieger2017, herrerias, whitlock}. Bringing these mature fields closer together via simplified models could provide a clearer understanding of biological processes in marine environments, as well as the effect of global climate change on our oceans and atmosphere \cite{behrenfeld}. 

Consider a population of passive organisms that are restricted to live at a specific depth. When parcels of incompressible water carrying organisms come together in a convergence zone, or water from deeper ocean layers rises towards the surface in an upwelling zone, organisms experience an effectively compressible velocity field. Other possible sources of effective compressibility include large Stokes numbers and gyrotaxis \cite{pigolotti2012}. 

We report here results from a one-dimensional agent-based stochastic model of two-species competition that allows for non-uniform occupation in continuous space, coupled to a compressible flow. Though fundamentally a three-dimensional problem, a one-dimensional approach has proven fruitful in the past. The analysis and computational overhead are significantly simplified, and trends observed in one dimension often hold in higher dimensions as well \cite{benzi, pigolotti2012}. We first study a sinusoidal velocity field to understand the effect of a stationary source (positive slope zero crossing) and sink (negative slope zero crossing) pair, and then apply this understanding to a shell model of turbulence. 

We find that Kimura's famous formula for the fixation probability in well-mixed systems \cite{kimura}, and spatially extended systems with only diffusive motion \cite{maruyama1974}, breaks down for even weakly compressible flows, dramatically increasing the influence of organisms born near source regions and lowering the overall probability of fixation for a given selective advantage, initial fraction, and system size. We explore this deviation, and are able to predict the scaling behaviors observed in simulations with simple theoretical models. We believe that the source-oriented view presented here provides a promising framework for predicting ecosystem outcomes in the presence of species mutation and invasion. Our results suggest that the effect of vertical velocities must be treated with care, even in simple models. 

\section{Model for Spatial Population Genetics with Compressible Advection}
The coarse-grained dynamics in the deterministic limit of our two-species stochastic model in one dimension are described by two coupled partial differential equations (see Ref. \cite{pigolotti2012} and SI Parts A and B):
\begin{equation}
\frac{\partial c}{\partial t} + \partial_x (u c) = D \partial_x^2 c + \mu c (1- c),
\label{FKPP} 
\end{equation}
\begin{equation}
\frac {\partial f }{\partial t} + u \partial_x f = D \partial_x^2 f + \frac{2 D}{c} \partial_x f \partial_x c + s c \mu f (1-f).
\label{genFKPP}
\end{equation}
where $c(x)$ is the fraction of the no-flow carrying capacity at position $x$, $f(x)$ is the fraction of organisms of a given species at $x$, $u(x)$ is a compressible velocity field, $D$ is the diffusion constant, $\mu$ is the growth rate when either species is dilute, and $s$ is the selective advantage of one species over the other when the system is near its no-flow carrying capacity, as defined by the microscopic rates given in SI Part A. 

In the limit of incompressible flow (i.e. when $u(x)$=const., as must be the case in one dimension), Equation \ref{FKPP} is the Fisher equation and admits traveling-wave solutions with speed $v_p=2 \sqrt{D \mu} + u$. Equation \ref{genFKPP}, describing the more complicated genetic dynamics, also reduces to a similar form in the limit $c(x) \to 1$, with a genetic wavefront speed of $v_g=2 \sqrt{D \mu s} + u$. We assume the selective advantage is small ($s <<1$), allowing us to define three parameter regimes, which will also be important in the compressible case, using the local value of $u(x)$:
\begin{enumerate}
\item $|u| > 2 \sqrt{D \mu}$: An opposing flow can arrest a Fisher population wave. Compressible flows of this magnitude can localize the population, for example, near a sink. Population dynamics in this localized regime have been studied in one and two dimensions \cite{benzi, perlekar2010, pigolotti2013}. \label{strong}
\item $2 \sqrt{D \mu s} < |u| < 2 \sqrt{D \mu}$: An opposing flow can arrest fragile Fisher genetic waves, but can only slow down the more robust Fisher population waves. A compressible flow near a one-dimensional sink is not able to create a localized steady state population structure, but can nevertheless localize genetic boundaries.  \label{extended}
\item $|u| < 2 \sqrt{D \mu s}$: An opposing flow is so weak that it can arrest neither genetic nor population waves. \label{weak}
\end{enumerate}

Here, we examine regimes \ref{extended} and \ref{weak}, which, to our knowledge, have not yet been systematically explored and have clear biological relevance. While some vertical velocities at strong upwellings certainly satisfy condition \ref{strong} (for example, we can estimate $D \approx 10^{-13} \text{m$^2$/s}$ by assuming unflagellated microorganisms with a Stokes-Einstein diffusivity $D=k_B T/6 \pi \eta R$, $R \approx 1 \mu \text{m}$, and assume $\mu$ to be $1 \text{ day}^{-1}$, giving us $2 \sqrt{D \mu} \approx 2 \times 10^{-4}$ m/day, whereas vertical velocities contribute roughly $|u|_{max} \approx 5 \text{ m/day}$ \cite{jacox, bravo, pollard1992, moore}), strongly localized structures at upwellings/convergence zones are not the only interesting situation. Conditions \ref{extended} and \ref{weak} describe weaker upwelling/convergence events, other sources of compressibility and/or a strong upwelling event in a population that has a greater effective diffusivity due to active flagella \cite{pigolotti2012}.
 In these cases, the steady state concentration profile is almost identical to the no-flow case, making them more theoretically tractable. Note that the presence of noise will change the location of the boundaries between the behaviors. For a detailed treatment of noisy Fisher waves, see Ref. \cite{hallatschek, doering}.

We further focus our investigation on fixation probabilities, a central topic in population genetics describing the stochastic process by which one species outcompetes another. The fixation probability for a species with selective advantage $s$ in a population of size $N$ that makes up an initial fraction $f$ of all organisms in the absence of advection and mutation is given by Kimura's formula, which neglects terms of order $s/N$ \cite{kimura}.
\begin{equation}
P_{fix}=\frac{1-\exp(-s N f)}{1-\exp(- s N)}.
\label{kim}
\end{equation}
This formula was first derived for the well-mixed case. However, it can be extended to one-dimensional systems with diffusive motion using an argument inspired by Maruyama \cite{maruyama1974} (see SI Part C), and is confirmed by our simulations. This insensitivity to spatial dimension makes the fixation probability an interesting object of study in the presence of fluid flows, because it allows us to isolate the effect of advection from that of diffusion. In contrast, fixation times, studied in the presence of advection by Pigolotti et al. \cite{pigolotti2012}, depend sensitively on the spatial structure of the system, and the well-mixed result for fixation times does not hold in the presence of diffusion. 

In this paper, we study the behavior of fixation probabilities for weakly compressible flows. As is often the case in population genetics \cite{hartlclark}, we characterize the altered fixation probabilities by an effective population size. For small $s$, the derivative of Equation \ref{kim}, with respect to $s$, is
\begin{equation}
\frac{d P_{fix}}{d s} = \frac{1}{2} N f(1-f) + \mathcal{O}(s).
\label{dkim}
\end{equation}
The slope at $s=0$ in the well-mixed case is proportional to the probability of a competitive encounter between the two different species. With advection, we can interpret the slope as measuring only encounters that have an impact on the future of the system, occurring for an effective population size such that $N \to N_{eff}$ in Equation \ref{dkim}. By avoiding spatially localized structures associated with strong flows, where $|u|> 2 \sqrt{D \mu}$, we can approximate the overall density of organisms as constant. This assumption greatly simplifies the determination of the effective population size, reducing it to an effective length scale. As we shall see, the result is to replace $N$ by $N_{eff}$ in the small-$s$ limit appropriate to Equation \ref{dkim}, and to replace $N$ by $N_g(s)$ in the large-$s$ limit, where the population genetics are dominated by Fisher genetic waves. 

\section{Sine Wave Flows}
To better understand the impact of sources (positive slope zero crossings) and sinks (negative slope zero crossings) on fixation prior to tackling time-dependent turbulence, we first study a steady sine wave flow given by
\begin{equation}
u(x)=A_0\sin(x-\pi/2),
\end{equation}  
in a domain of size $2 \pi$ with periodic boundary conditions. The source and sink associated with this velocity profile have a characteristic time, given by $\tau_s=1/A_0$, the inverse gradient of the velocity field at the zero crossings. Organisms are more likely to die near the sink at $3 \pi/2$, where there is a constant influx of organisms, and more likely to flourish near the source at $\pi/2$. These effects violate the conditions necessary for Kimura's formula to hold in one dimension (see SI Part C, \cite{maruyama1974}), and give an advantage to the source population relative to the sink population, even if there is no microscopic selective advantage involved. To determine the effective population size, we need to characterize the width of the advantageous source region.

\begin{figure}
\begin{center}
\includegraphics[scale=0.22]{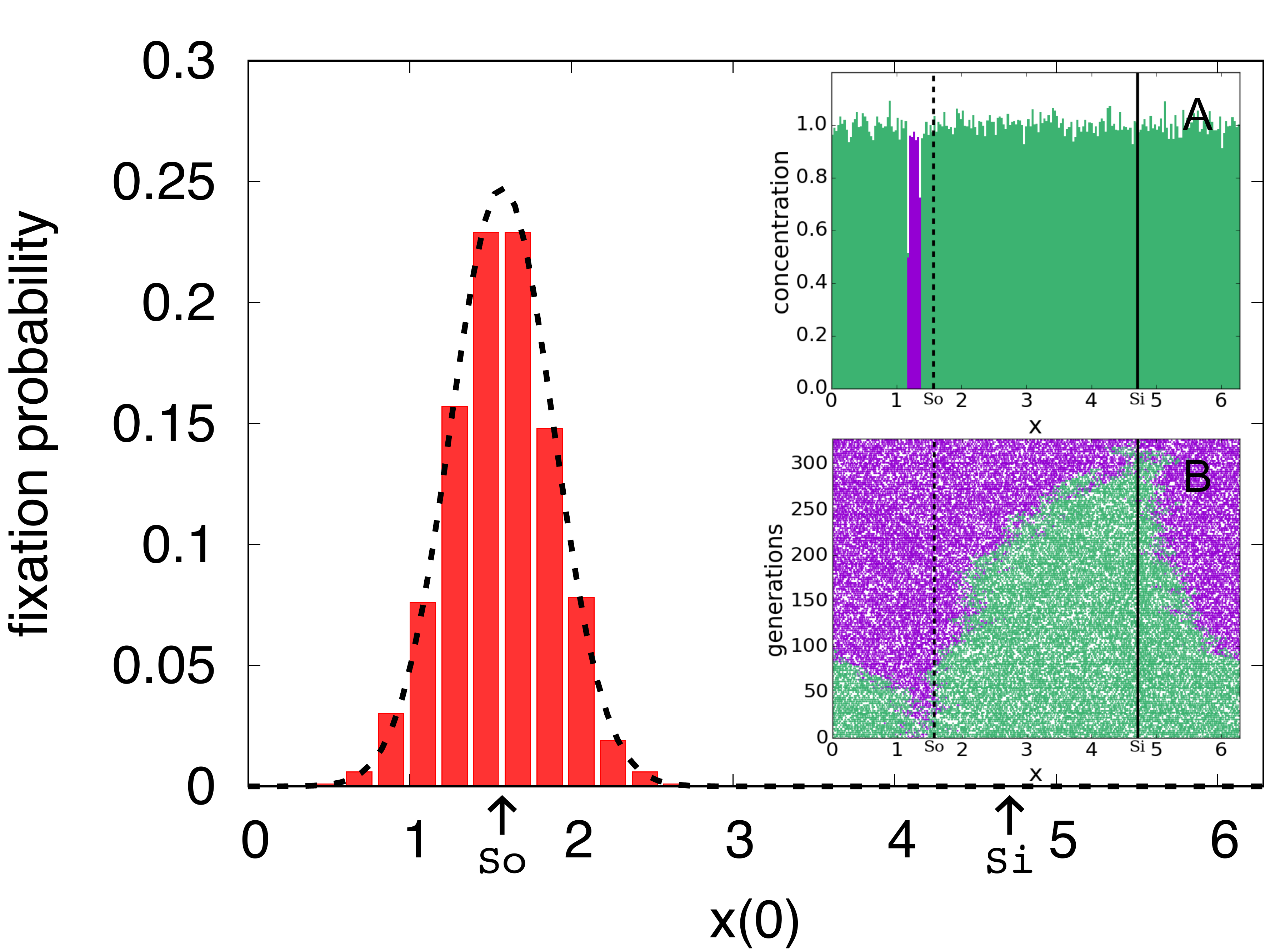}
\end{center}
\caption{{\label{rwfig} The fixation probability of an initially localized purple species in a background of green, obtained from 1000 independent realizations of an agent-based simulation of two neutral populations, as a function of the initial position of the purple distribution. Arrows indicate the locations of the source ($So$) and sink ($Si$). Populations that begin near the source in the sine wave flow are more likely to fix. The dotted line shows agreement with the random walk model, Equation \ref{rwers}, with no fitting parameter. The flow for this plot is $u(x)=0.05 \sin(x-\pi/2)$. (inset A): One possible initial condition, with $1/32$ of the domain holding only purple organisms, and the rest filled with green organisms. (inset B): One realization of a fixation event corresponding to the initial condition in (A). The dashed line shows the source at $\pi/2$ and the solid line shows the sink at $3 \pi/2$.
}}
\end{figure}

\subsection{Quasi-neutral Competitions with Sine Wave Flows}
For the case of zero selective advantage, we can analytically compute an approximate fixation probability as a function of position using a random walk model. As shown in Figure \ref{rwfig}, simulations are initialized so that only a small, localized window (length $\pi/16$, or $\approx 3 \%$ of the interval [0, $2 \pi$] in Figure \ref{rwfig}) contains all of one species. The fixation probability is then measured as a function of the location of this window. By treating these genetic boundaries as random walkers biased by the flow, we find in the limit of small $\Delta$,
\begin{equation}
P_{fix}(x, \Delta, A_0) \approx \Delta\cdot \mathcal{N}(x | \pi/2, D/A_0),
\label{rwers}
\end{equation}
where $\Delta$ is the width of the spatial window, $x$ is the leftmost genetic boundary, and $\mathcal{N}(x | \pi/2, D/A_0)$ is the normalized probability density function at $x$, given by a Gaussian with mean $\pi/2$ and variance $D/A_0$.
Details are presented in SI Part D. 

This function defines a length scale for our source, $l_s=\sqrt{D/A_0}$, given by the balance between diffusion and the advecting velocity field. Any organism that can diffuse to the source in a time faster than the source time $\tau_s = l_s^2/D$ is relevant to the genetic future of this sine wave system. The farther away an organism is from the source, the stronger the outward velocity field it experiences. Any organism that moves significantly farther than $l_s$ from the source is unlikely to be able to return, and has a negligible chance of fixation as it is drawn into the sink.

\begin{figure}
\begin{center}
\includegraphics[scale=0.22]{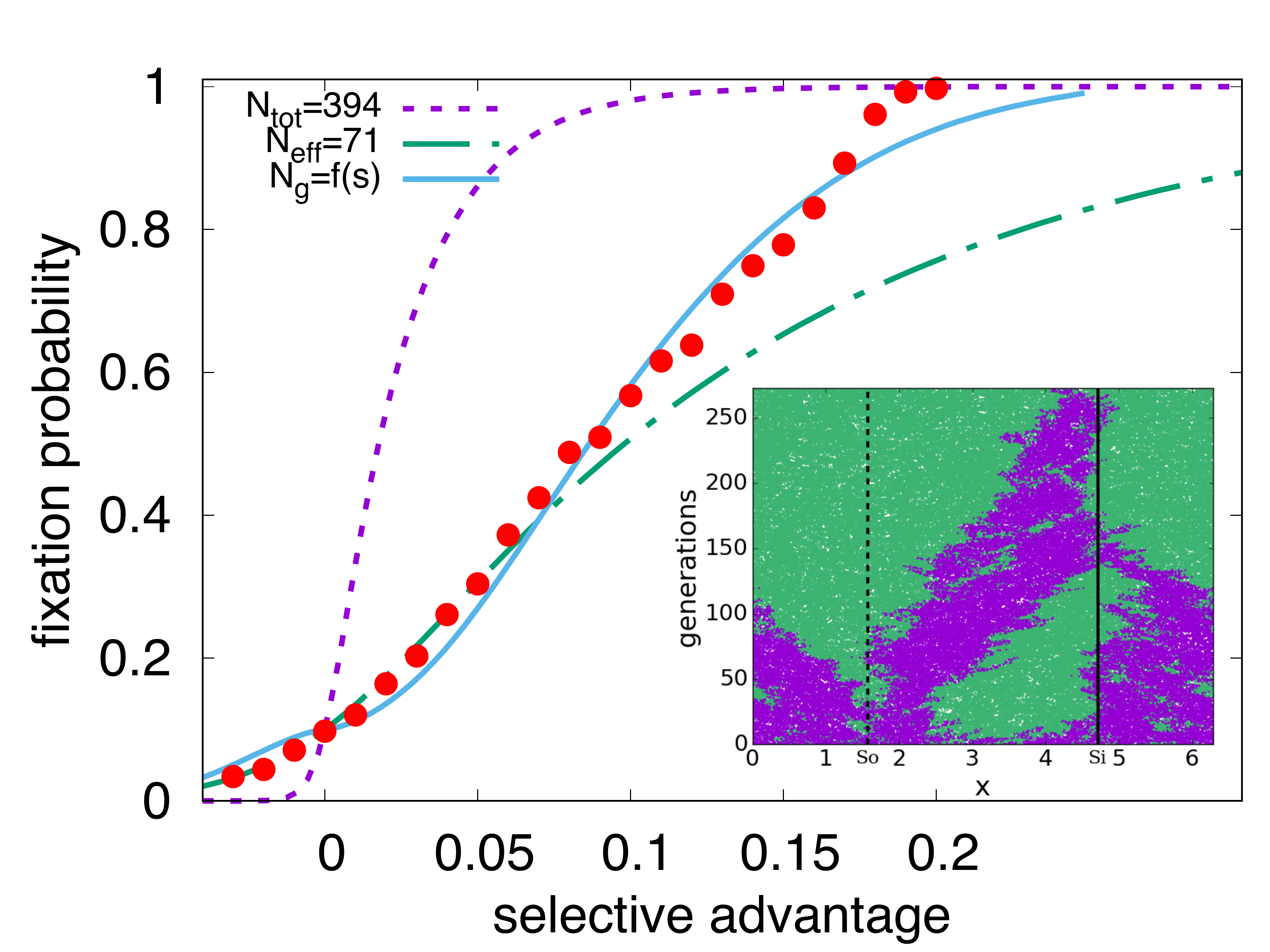}\\
\caption{{\label{sinecross1} The fixation probability (red circles) for an initial condition with a random 10/90 mixture of two species with $u(x)=0.05 \sin(x-\pi/2)$, varying the selective advantage ($P_{fix}(s=0)=0.1$, since all organisms are equally likely to take over the population in this limit).  The purple line (short dashes) shows Equation \ref{kim}, the Kimura no-flow result, for the measured number of organisms in the simulation, $N_{tot}=394$. The green line (short and long dashes) gives the prediction of Equation \ref{ncut}, valid in the limit of small $s$ with $N_{eff}=71$. The blue line (solid) gives the prediction of Equation \ref{ng} for the large $s$ case, with a selective advantage-dependent effective population size. Error bars inferred from 2000 independent simulations are too small to be visible. Inset: A single realization for a 50/50 mixture of two neutral species (thus, $P_{fix}(s=0)=0.5$).  The dashed line marks the position of the source, $So$, and the solid line marks the position of the sink, $Si$.  Note that the genetic interfaces between purple and green tend to annihilate in the sink at $3\pi/2$. The total concentration of organisms remains approximately uniform.
}}
\end{center}
\end{figure}

We hypothesize that for zero or small selective advantage the effective population size scales as the source length scale, times the density of organisms, $\rho_0$, which is a constant well-approximated by the no-flow limit since we are not in the regime of spatial localization. Based on these considerations, we conclude that
\begin{equation}
N_{eff} =B_1 \rho_0 \sqrt{\frac{D}{A_0}},
\label{ncut}
\end{equation}
where $B_1$ is a constant of order unity.
We see from Figure \ref{sinecross1} that the fixation probability is dramatically different from the case without an external velocity field, and is well described by these considerations for small $s$, with $B_1=3.5$. The large difference between $N_{eff}$ ($N_{eff}=71$) and the total number of organisms in the system ($N_{tot}=394$) is striking considering that the presence of the flow cannot be easily detected in the snapshots of the organismic density alone without observing genetic interfaces. 

\begin{figure}
\includegraphics[scale=0.70]{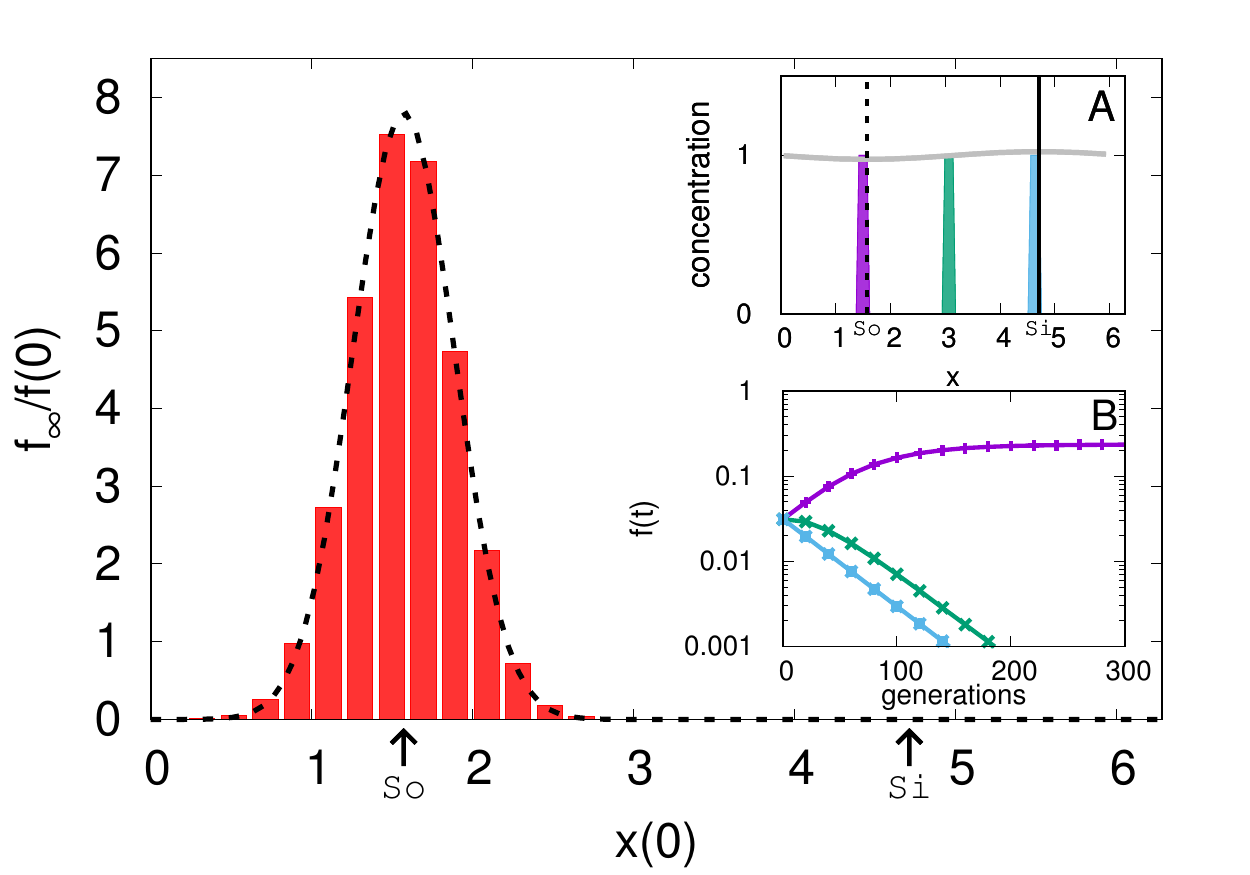}
\caption{{\label{edet}  The final fraction of the initially localized species, $f_\infty= \lim\limits_{t \to \infty} f(t)$, divided by its initial fraction, as a function of its initial location. Populations that begin near the source, $So$ ($x=\pi/2$) grow relative to their initial condition, and populations that begin far from the source shrink. Fit is given by a Gaussian of variance $l_s^2=D/A_0$.  Here $u(x)=0.05 \sin(x-\pi/2)$, as in Figure \ref{rwfig}. (inset A): The grey line shows the approximately uniform steady state total concentration of organisms (the sum of the concentrations of both species, $c(x,t)$). The colored boxes show the initial spatial distribution of one of the species for three different simulations, relative to the source (dashed line) and sink (solid line). (inset B): The total fraction of the initially localized species in time, with the colors corresponding to the initial conditions in (A). The leftmost initial population prospers, while the middle and rightmost initial populations fall off rapidly.
}}
\end{figure}

Although a stochastic model was the key to predicting the magnitude of the fixation probability given by Equation \ref{rwers}, the Gaussian enhancement of the source region is also evident in a purely deterministic model. Deterministic simulations of the neutral case, which directly solve Equations \ref{FKPP} and \ref{genFKPP} for $s=0$ without number fluctuations, are shown in Figure \ref{edet}. Although these simulations, which are equivalent to the agent-based model in the limit $N \to \infty$, cannot directly observe fixation, we nevertheless see that a small population of one species initially localized close to the source (inset A) grows in size until it reaches a steady state population (inset B) more than five times its original size. In contrast, a population starting near the sink shrinks to values that are less than $1/N$, approximating extinction in a simulation with discrete organisms.  A Gaussian with standard deviation $l_s$ centered on the source provides a good fit for Figure \ref{edet}, just as in Figure \ref{rwfig}.

\subsection{Strong Selective Advantage with Sine Wave Flows}
When the selective advantage is sufficiently high (although still small compared to 1), it becomes a significant term in Equation \ref{genFKPP} and plays a role in setting the source length. We must now balance the effects of selection, diffusion, and the velocity.

We can do this using the framework of Fisher genetic waves. Consider a species with a significant selective advantage that is sharply localized within a background population, in a system with a sine wave flow with an amplitude that nevertheless satisfies $A_0> 2\sqrt{D \mu s}$. This initial condition will typically produce two Fisher genetic wavefronts traveling in opposite directions. If the initial population starts on a source, both wavefronts will be supported by the external flow as they move across the system. However, if the initial population starts on a sink, both wavefronts will face an opposing flow. 

There is a window of initial conditions around the source at $x=x_s=\pi/2$, defined by $u(x)= A_0 \sin(x-\pi/2)=\pm 2 \sqrt{D \mu s}$, where an initial population can produce Fisher genetic waves that deterministically travel across the entire system. Within this window, the flow velocity is small enough that wavefronts can reach the source even if they do not start there. Outside of this window, Fisher genetic waves cannot reach the favorable source region.

Upon treating Fisher waves crossing the system as a proxy for fixation, this argument suggests another way to define an effective population size in the Kimura formula: the number of organisms around the source within the spatial window with boundaries given by $u(x)= \pm 2 \sqrt{D \mu s} $. 

When we solve for this window, we include a fitting parameter, $B_2$, because the traveling wavefront solution only occurs for special initial conditions. The window is then given by half-width $\delta(s)$ such that
\begin{equation}
A_0 \sin(\delta)= B_2 (2 \sqrt{D \mu s}).
\end{equation}

After solving for $\delta(s)$, we can use the approximation of constant density to write the number of organisms in our genetic wave-defined source population as
\begin{equation}
N_g= 2 \rho_0 \delta(s).
\label{ng}
\end{equation}

Note that our effective population size $N_g$ now has $s$ dependence. We find that $B_2=0.5$ provides good agreement in Figure \ref{sinecross1}. At even higher values of selective advantage, where $A_0$ is no longer greater than $2 \sqrt{D \mu s}$ we observe a crossover to well-mixed behavior as expected.

\begin{figure}
\includegraphics[scale=0.68]{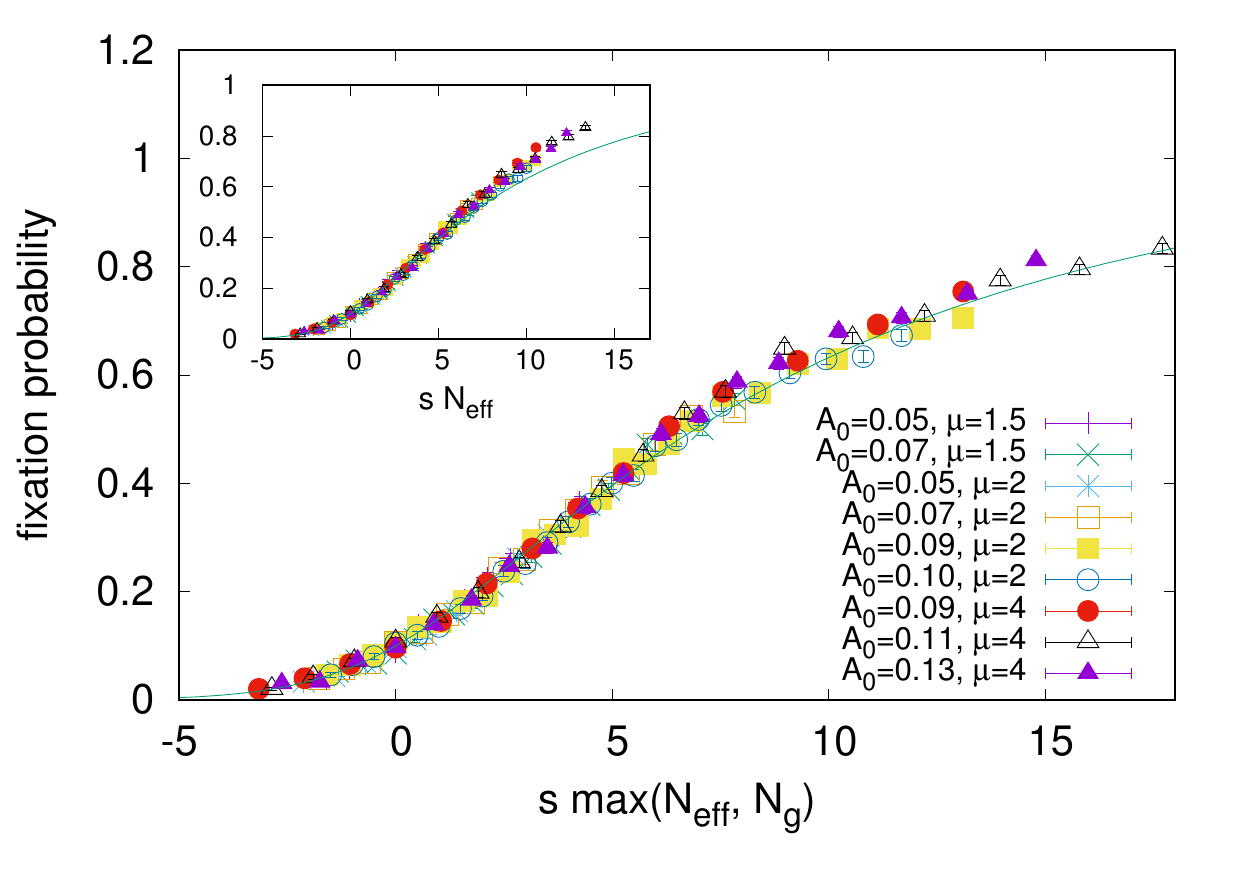}
\caption{{\label{sinescale} 
When we plot the measured fixation probabilities against $s$ times the larger effective population size of our two approximations, Equations \ref{ncut} and \ref{ng}, our data collapse to a master curve. The agreement with our sine wave simulations suggests that we have identified how the fixation varies with $A_0, \mu,$ and $s$ for this parameter regime. Error bars associated with 2000 independent realizations are shown. $A_0 > 2 \sqrt{D \mu s}$ for all data points shown, avoiding the crossover to well-mixed behavior seen at high $s$ in Figure \ref{sinecross1}. Inset: When plotted as a function of the quasi-neutral theory only, Equation \ref{ncut}, we see a departure from the theory at high values of $s$, which motivates the high $s$ approximation given by Equation \ref{ng}. 
}}
\end{figure}

Upon combining these two arguments by using the largest effective population size, $\max(N_{eff}, N_g)$, in the original Kimura formula, Equation \ref{kim}, we can explain the fixation probability for a weak steady sine wave, as shown in Figure \ref{sinescale}. 

\section{Turbulent Advection}
To generate a one-dimensional chaotic signal with multiscale correlations similar to turbulent flow, we use a well-established set of ordinary differential equations called a shell model \cite{biferale}. For specific details, see Ref. \cite{benzi} and SI Part E. For the steady sine wave flows studied in the previous section, we saw that the presence of a source can dominate fixation events. It seems reasonable to conjecture that transient, time-dependent sources have a similarly important effect in chaotic turbulent flows, and that we can still characterize the system with an effective source dimension and an associated effective population size. Remarkably, we can understand much of the fixation probability with generalizations of the simple theoretical arguments we applied to the stationary sine wave case.

\subsection{Quasi-neutral Competitions with Turbulent Advection}
Even in the absence of selection, turbulent dynamics provide new factors to consider when determining the source length scale. The (multiple) sources in the flow can have long or short lifetimes. Their locations move intermittently, and their slopes vary with time. Depending on how the source times compare with the other timescales dictating organism motion, the source-enhanced population at a given time may or may not be able to move with the source and retain its competitive advantage. An example of competitive turbulent dynamics for the neutral case and some further discussion are given in SI Part F. Although it is clear that understanding the details of fixation probabilities in turbulence is complicated, we can nevertheless apply lessons from the sine wave case to determine scaling behavior. 

We assume that our effective population size is the sum of the effective population sizes corresponding to each source, averaged over time.  Longer-lived sources thus contribute more to the average and have a greater effect on the effective population size. As in the case of the weakly compressible sine wave, a source's effective population size should be able to be represented as the density of organisms times a characteristic length. It is not obvious how to define a length for an arbitrary source $i$, but we can easily define a characteristic source time, $\tau_i$, by taking the reciprocal of the slope of the positive zero crossing ($\frac{d u(x)}{d x}\big|_{x=x_i}^{-1}$). Then, we construct a length by assuming the existence of a constant with units of velocity, $v_C$, that can depend on the diffusion constant and some details of the dynamics, but does not depend on the root mean-square velocity. We can think of $v_C$ as related to the speed with which domain boundaries explore the system.

Therefore, our estimate for $N_{eff}$ in turbulence is
\begin{equation}
N_{eff}= \rho_0 v_C \overline{\sum_i \tau_i}.
\label{neff1}
\end{equation}
Here, the sum is over all sources present at a given time, and the overbar indicates a time average.

Since the density of organisms, $\rho_0$, is approximately constant, it is proportional to $\mu$, the growth rate. Therefore, we can also understand Equation \ref{neff1} as a balance between the organism generation time, $\mu^{-1}$ and the source time, $\tau_i$.  If $\mu \tau_i$ is large, the generation time is short relative to the source time, and organisms can reproduce many times during the source time. These organisms and their offspring thus experience an enhanced fixation probability, and such sources will give a large contribution to $N_{eff}$. However, if $\mu \tau_i$ is small, few organisms are affected, and the contribution to $N_{eff}$ will be small. 

We now make the approximation
\begin{equation}
\overline{\sum_i \tau_i} \approx \overline{n_s \tau_s},
\end{equation}
where $n_s$ is the number of sources and $\tau_s=\Big<\Big(\frac{\partial u(x,t)}{\partial x}\Big)^{-2}\Big>^{1/2}$ is the root mean-square reciprocal velocity gradient.

We expect that, for a given value of the root mean-square velocity, $u_{rms}$, the total number of zero crossings, $2 n_s$, scales with its gradient. As the root mean-square velocity increases, we expect the number of zero crossings to decrease. These considerations lead to the conjecture
\begin{equation}
n_s \sim \frac{L}{u_{rms}}\Big<\Big(\frac{\partial u(x,t)}{\partial x}\Big)^2\Big>^{1/2}.
\end{equation}
This is known to be true for Gaussian processes in one dimension \cite{rice}, and we have checked it explicitly via simulations with our shell model. A similar relation has been found experimentally in measurements of turbulent flows, where the number of nodes is proportional to the inverse of the Taylor microscale \cite{kaila}. 

We note that $n_s$ and $\tau_s$ are instantaneously strongly fluctuating quantities, and we calculate the dependence of their product on Reynolds number in SI Part G as $n_s \tau_s \sim \text{Re}^{0.08}$. Since this dependence is very weak, we neglect it and make a mean field approximation to find $\overline{n_s \tau_s} \approx L/u_{rms}$, where $u_{rms}$ is now time averaged using the harmonic mean. 

Upon combining these arguments, and absorbing $v_C$ into the constant $B_3$, we obtain
\begin{equation}
N_{eff}=B_3 \rho_0\overline{n_s \tau_s} = \frac{B_3 \rho_0 L}{u_{rms}}.
\label{turbeff2}
\end{equation}

\begin{figure}
\begin{center}
\includegraphics[scale=0.22]{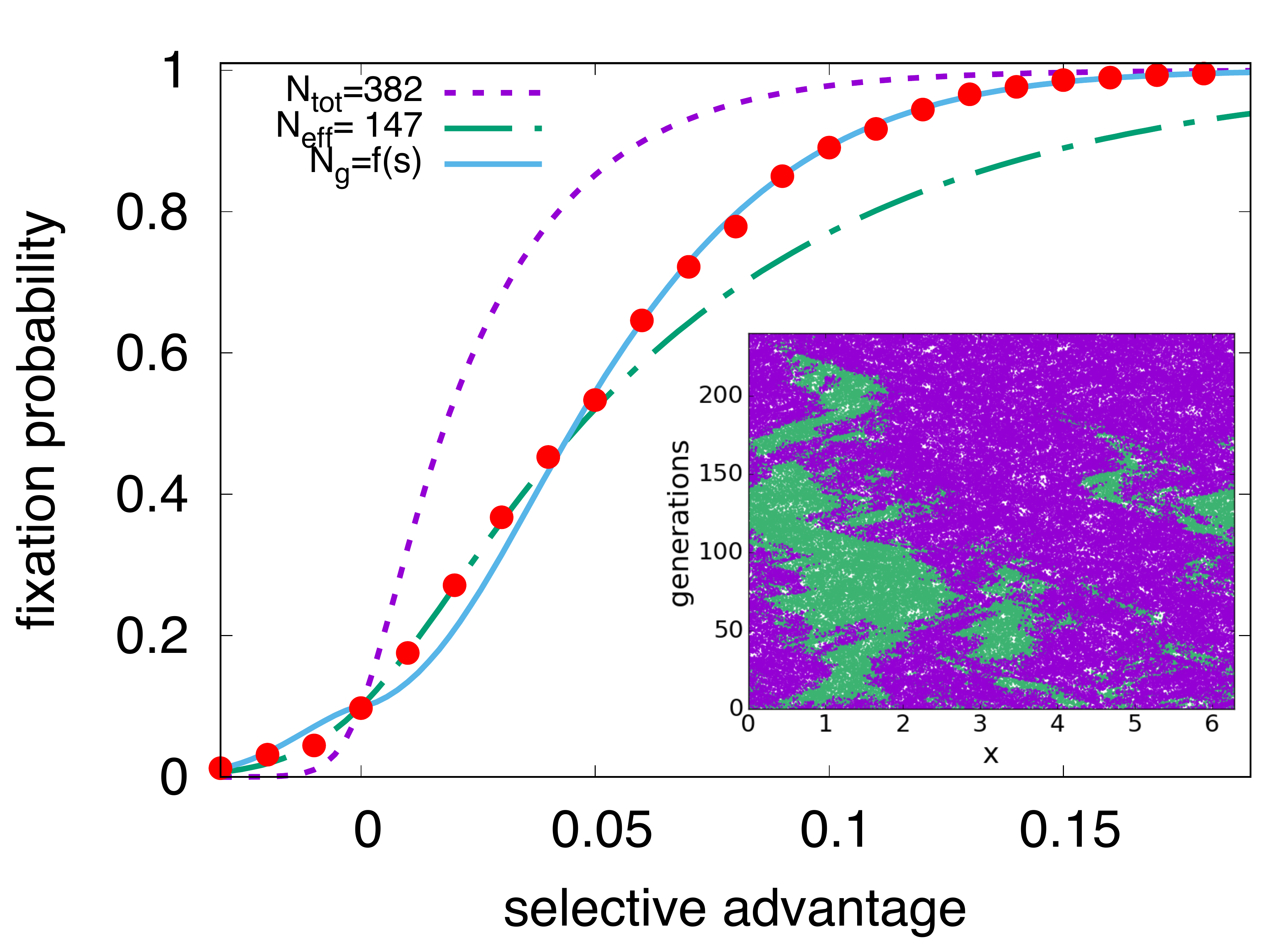}
\caption{{\label{extturb2}   The fixation probability (red circles) for an initial condition with a random 10/90 mixture of two species advected by a shell model with amplitude $A_0=0.3$ and growth rate $\mu=2$, for variable selective advantage ($P_{fix}(s=0)=0.1$).  The purple line (short dashes) shows Equation \ref{kim}, the no-flow result, for the measured number of organisms in the simulation, $N_{tot}=382$. The green line (short and long dashes) gives the prediction of Equation \ref{ncut}, valid in the limit of small $s$ with $N_{eff}=147$. The blue line (solid) gives the prediction of Equation \ref{ng} for the large $s$ case, with a selective advantage-dependent population size. Error bars inferred from 2000 independent simulations are too small to be visible. Inset: A single realization for a 50/50 mixture of two neutral species ($P_{fix}(s=0)=0.5$).  
}}
\end{center}
\end{figure}

Our simulations support this form of $N_{eff}$ as shown in Figures \ref{extturb2} and \ref{turbscale} with the constant (units of velocity to account for unknown $v_C$ factor) $B_3=0.031$.

\subsection{Strong Selective Advantage with Turbulent Advection}
As in the sine wave case, we expect a critical value of $s$ beyond which selection must be taken into account in the source size calculation. Unlike the quasi-neutral competitions in turbulence, however, we now have an obvious choice for a velocity that can be used to form a length scale-- the Fisher genetic wavefront speed. As before, we include a dimensionless fitting parameter, $B_4$.
Our estimate for $N_g$, the effective population size associated with genetic waves, is
\begin{equation}
N_g = B_4 \rho_0(2 \sqrt{D \mu s})\overline{\sum_i \tau_i},
\label{neff2}
\end{equation}
where $\tau_i$ is the characteristic time of source $i$ and the overbar indicates a time average. 

Equation \ref{neff2} is also the simplest generalization of Equation \ref{ng}, obtained by expanding the sine function to linear order in a Taylor series close to each source.

\begin{figure}
\includegraphics[scale=0.68]{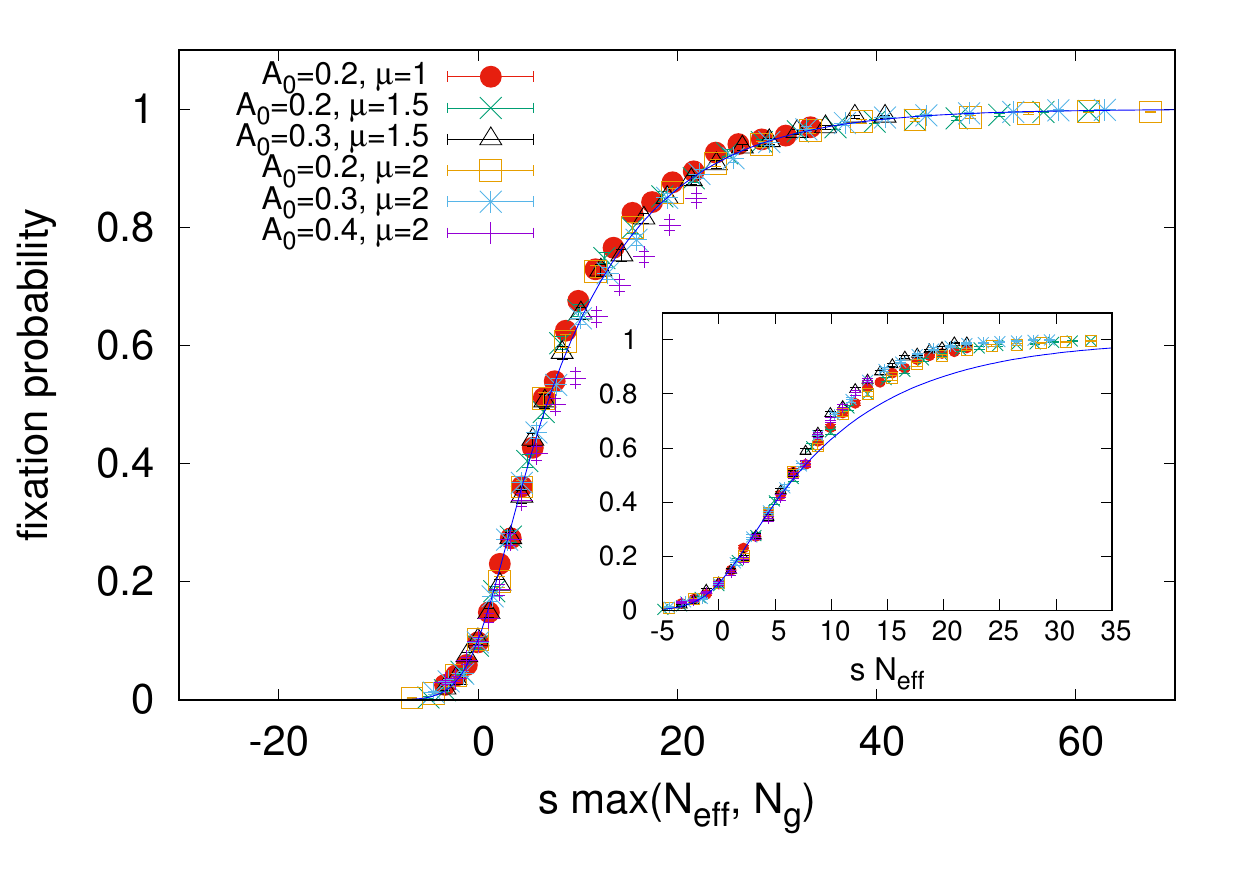}
\caption{{\label{turbscale}  When we plot the measured fixation probabilities against $s$ times the larger effective population size of our two approximations, Equations \ref{turbeff2} and \ref{turbng2}, our data collapse to a master curve. The agreement with our shell model simulations suggests that we have identified how the fixation varies with $A_0, \mu,$ and $s$ for this parameter regime. Error bars associated with 2000 independent realizations are shown.  $N_{tot} \approx N_0$ for all data points shown, avoiding the crossover to localized behavior. These parameters ensure that approximate spatial uniformity is maintained. Inset: When plotted as a function of the quasi-neutral theory only, Equation \ref{turbeff2}, we see a departure from the theory at high values of $s$, which motivates the high $s$ approximation given by Equation \ref{turbng2}. 
}}
\end{figure}

Upon estimating $\overline{\sum_i \tau_i}$ as before, we obtain
\begin{equation}
N_g = B_4 \rho_0 L \frac{2 \sqrt{D \mu s}}{u_{rms}}.
\label{turbng2}
\end{equation}
This estimate, when $N \to N_g$ in the Kimura formula, Equation \ref{kim}, shows good agreement with Figures \ref{extturb2} and \ref{turbscale}, with $B_4=0.747$.
As before, we combine the arguments behind Equations \ref{turbeff2} and \ref{turbng2} by taking the largest effective population size, $\max(N_{eff}, N_g)$, in Figure \ref{turbscale}. 

\section*{Discussion}
For both simple (i.e. sine wave) and turbulent compressible flows for which the population density is approximately uniform, we have shown that fixation probabilities are controlled by an effective population size smaller than the total number of organisms in the system in a previously undescribed and biologically relevant region of parameter space. This reduction in the effective population size creates a significant reduction in the fixation probability as a function of selective advantage overall, but a greatly enhanced fixation probability for organisms fortunate enough to be born near sources, even those that are very weak. In the ocean, source regions can be associated with upwellings, if we assume organisms are restricted to live at a certain depth, for example. Our results suggest that the genetic compositions of these regions may have a controlling effect on the genetics of a much greater domain.

Furthermore, we have shown the reduced fixation probabilities can be explained by simple theoretical arguments and can be explored with both agent-based and relatively inexpensive deterministic simulations. 

Investigations are currently underway to examine to what extent these results hold in two dimensions and for more realistic flows. Deviations of Kimura's formula in the case of strongly compressible turbulence in one dimension (i.e. flows that produce spatial localization) have also been observed in simulations, a problem closely related to gene surfing \cite{hallatschek2008}. 

\section*{Acknowledgments}
We are grateful for the assistance provided by Pinaki Kumar and Francesca Tesser. We thank Amala Mahadevan, Mara Freilich, and Luca Biferale for useful discussions. Work by AP and DRN was supported by the National Science Foundation, through grants DMR-1608501 and via the Harvard Materials Science Research and Engineering Center via grant DMR-1435999. Work by F.T. was partially supported by the Nederlandse Organisatie voor Wetenschappelijk Onderzoek I (NWO-I), The Netherlands.

%

\end{document}


\title{Supporting Information (SI)}
\author{Abigail Plummer, Roberto Benzi, David R. Nelson, Federico Toschi}

\maketitle
\vspace{-3em}
\tableofcontents{}
\vspace{-1em}

\subsection{Agent-based Simulations with Advection and Diffusion}
\label{micro}
For a well-mixed system, our stochastic agent-based model for two species, $A$ and $B$, uses reaction rates:
\begin{align}
\text{birth of $A$:} \hspace{2em }  A  &\xrightarrow{} A + A  \hspace{0.2 in} \text{at rate $\mu$}, \label{rules1} \\
\text{death of $A$:}\hspace{2em } A  &\xrightarrow{} \emptyset  \hspace{0.2 in} \text{at rate $\gamma (N_A - 1) + \lambda_{AB} N_B$},\\
\text{birth of $B$:}\hspace{2em }B  &\xrightarrow{} B + B  \hspace{0.2 in} \text{at rate $\mu$}, \\
\text{death of $B$:}\hspace{2em }B  &\xrightarrow{} \emptyset  \hspace{0.2 in} \text{at rate $\gamma (N_B - 1) + \lambda_{BA} N_A$}, 
\label{rules4}
\end{align}
where $N_A$ and $N_B$ are the number of organisms of species $A$ and $B$ respectively.

We assume that birth and intraspecies death processes occur at the same rates, $\mu$ and $\gamma$, for both species, making the well-mixed carrying capacity of each species in isolation identical. Interspecies death rates can vary, however, allowing the model to capture selective advantage, mutualism, and competitive exclusion \cite{pigolotti2013}. We are interested in the case of competition in the presence of selective advantage, and take, for simplicity, $\gamma$, the intraspecies death rate, to be equal to the average of $\lambda_{AB}$ and $\lambda_{BA}$. The deviation from this average value, normalized by $\gamma$, is a measure of the selective advantage, $s$. 
\begin{equation}
s=1-\frac{\lambda_{AB}}{\gamma}= \frac{\lambda_{BA}}{\gamma}-1.
\label{sel}
\end{equation}

Another common choice for selective advantage is to let $\mu \to \mu (1+s)$ for one of the species, modeling a faster growth rate. This choice represents selective advantage in dilute conditions, while Equation \ref{sel} gives selective advantage under crowded conditions \cite{chotibut}. Our definition simplifies our analysis because it leads to a stationary average total population size throughout the fixation process. 

We extend our model to one dimension by dividing our spatial domain into intervals, and only allow organisms within the same interval to contribute to the death rates as defined above. Advection and diffusion are incorporated by allowing organism $i$ at position $x_i$ to move at each time step according to
\begin{equation}
\Delta x_i = u(x_i, t) \Delta t + \sqrt{2 D  \Delta t} \Gamma (t) ,
\end{equation}
where $D$ is the diffusion constant, $u(x, t)$ is the velocity field, and $\Gamma(t)$ is a normally distributed random variable with zero mean and unit variance.  

\subsection{Macroscopic Equations}
\label{macro}
Given the microscopic reaction rates in the previous section, we can carry out a coarse-graining procedure using a Kramers-Moyal expansion for a well-mixed system \cite{pigolotti2012, risken, gardiner}. The Fokker-Planck equation, in terms of Kramers-Moyal expansion coefficients $C_i^{(1)}(\mathbf{y})$ and  $C_{ij}^{(2)}(\mathbf{y})$, is
\begin{equation}
\partial_t P(\mathbf{y}, t)= - \sum_i \partial_i [C^{(1)}_i(\mathbf{y})P(\mathbf{y},t)] + \frac{1}{2} \sum_{i,j} \partial_i \partial_j [C^{(2)}_{ij}(\mathbf{y}) P(\mathbf{y},t)],
\label{fp}
\end{equation}
with the sum over species, and $P(\mathbf{y},t)$ is the probability of being in the state $\mathbf{y}=(a, b, ...)$, a vector formed from the number of organisms of each species, at time $t$.

For our set of reactions, we have two variables, $a$, the number of organisms of species $A$, and $b$, the number of organisms of species $B$. The Kramers-Moyal expansion coefficients for our system, using the microscopic parameters given in Section \ref{micro}, are
\begin{align}
C^{(1)}_a&= \mu a - \gamma a (a-1) - \lambda_{AB} a b \approx \mu a - \gamma a^2 - \gamma(1-s) a b, \\
C^{(1)}_b&= \mu b - \gamma b (b-1) - \lambda_{BA} a b \approx \mu b - \gamma b^2 - \gamma (1+s) a b ,\\
C^{(2)}_{aa}&= \mu a + \gamma a (a-1) + \lambda_{AB} a b \approx \mu a + \gamma a^2 + \gamma(1-s) a b ,\\
C^{(2)}_{bb}&= \mu b + \gamma b (b-1) + \lambda_{BA} a b \approx \mu b + \gamma b^2 + \gamma (1+s) a b, \\
C^{(2)}_{ab}&=C^{(2)}_{ba}=0,
\end{align}
where we have assumed $a, b >> 1$.

A set of stochastic differential equations associated with Equation \ref{fp} is
\begin{align}
\frac{d a}{d t} &= C^{(1)}_a + \sqrt{C^{(2)}_{aa}} \Gamma_a (t) , \label{langa}\\
\frac{d b}{d t} &= C^{(1)}_b + \sqrt{C^{(2)}_{bb}} \Gamma_b (t) , \label{langb}
\end{align}
where the $\{\Gamma_i(t)\}$ are delta correlated Gaussian random variables such that $\langle\Gamma_i(t)\Gamma_i(t^\prime)\rangle=\delta(t-t^\prime)$ and $\langle\Gamma_i(t) \Gamma_j(t)\rangle = \delta_{ij}$. Note that the noise is multiplicative.

We now restrict our attention to the deterministic parts of Equations \ref{langa} and \ref{langb} (i.e. we neglect the noise terms), as this is all we need for the Fisher wave analysis. We follow the procedure described in detail in \citet{pigolotti2012} to add advection and diffusion terms. The equations for $a(x,t)$ and $b(x,t)$, subject to a compressible flow field $u(x,t)$ in one dimension, become
\begin{align}
\frac{\partial a }{\partial t} + \partial_x (u a) &= D \partial_x^2 a + \mu a - \gamma a (a+b) + s \gamma a b , \label{deta}\\ 
\frac{\partial b }{\partial t} + \partial_x (u b) &= D \partial_x^2 b + \mu b - \gamma b (a+b) - s \gamma a b .\label{detb}
\end{align}

Upon changing variables to $f(x,t)= \frac{a}{a+b}$, the fraction of $A$ organisms, and $c(x,t)=\frac{(a+b)\gamma}{\mu}$, the total population as a fraction of the well-mixed carrying capacity, we recover Equations 1 and 2 of the main text,

\begin{equation}
\frac{\partial c}{\partial t} + \partial_x (u c) = D \partial_x^2 c + \mu c (1- c),
\label{FKPP} 
\end{equation}
\begin{equation}
\frac {\partial f }{\partial t} + u \partial_x f = D \partial_x^2 f + \frac{2 D}{c} \partial_x f \partial_x c + s c \mu f (1-f).
\label{genFKPP}
\end{equation}
Note that these equations only hold in the continuum limit. For small $a+b$, we expect these equations to break down. 

\subsection{Kimura's Formula, Extended to Higher Dimensions}

To derive Equation 3 in the main text, the fixation probability, we need to work with the Kolmogorov backward equation for our system in the long time limit with the final condition that no $B$ organisms remain. We start by deriving results for well-mixed organisms. At second order, the Kolmogorov backward equation for the well-mixed case is
\begin{equation}
\partial_{t^\prime} P(\mathbf{y^\prime}, t^\prime)=  \sum_i  C^{(1)}_i(\mathbf{y^\prime}) \partial_i P(\mathbf{y^\prime},t^\prime) + \frac{1}{2} \sum_{i,j}  C^{(2)}_{ij}(\mathbf{y^\prime}) \partial_i \partial_j P(\mathbf{y^\prime},t^\prime),
\label{bkwds}
\end{equation}
where $\mathbf{y}$, as before, is a vector of the number of organisms in the system, and $\mathbf{y^\prime}$ is the initial state of the system at time $t^\prime$.

We again change variables to $f= \frac{a}{a+b}$ and $c=\frac{(a+b)\gamma}{\mu}$. Since we are no longer neglecting noise, we first convert Equation \ref{bkwds} to a stochastic differential equation and then apply Ito's formula (\cite{gardiner}, page 93). Because we expect $N=\mu/\gamma$, the well-mixed carrying capacity, to be large and $s$ to be small, we can neglect terms that are multiplied by $\gamma s$.
This gives us the equations:
\begin{align}
\frac{d f}{dt} &= s c \mu f(1-f) + \sqrt{f(1-f)\gamma\Big( \frac{c+1}{c} \Big) } \Gamma_f(t) , \label{df}\\
\frac{d c}{dt} &= \mu c (1-c) + \sqrt{ \gamma c (c+1)} \Gamma_c (t) .
\label{dc}
\end{align}
Note that Equations \ref{df} and \ref{dc} differ in two important ways. First, if $s << 1$, Equation \ref{df} relaxes much more slowly than Equation \ref{dc}. Second, $f=1$ is an absorbing state, whereas $c$ can fluctuate freely around $c=1$. We will now decouple these two equations. In the absence of noise, $c$ is stationary at either $c=0$ or $c=1$, unstable and stable states respectively. To see if it is reasonable to approximate $c$ with $c(x, t)=1$, we linearize Equation \ref{dc} by setting $c=1+\varepsilon(t)$. 
\begin{equation}
\frac{d \varepsilon}{dt} =- \mu \varepsilon + \sqrt{\frac{2 \mu}{N}} \Gamma_c (t) .
\end{equation}
This Langevin equation describes an Ornstein-Uhlenbeck process, whose stationary solution is a Gaussian with mean 0 and variance $1/N$ \cite{gardiner}. Because of this $1/N$ scaling, it is reasonable to neglect fluctuations in c, corresponding to a large carrying capacity, and set $\varepsilon=0$, or $c=1$.  

Having eliminated Equation \ref{dc}, Equation \ref{df} becomes
\begin{equation}
\frac{d f}{dt} = s \mu f(1-f) + \sqrt{ \frac{2 \mu}{N} f(1-f)} \Gamma_f(t) ,
\end{equation}
which corresponds to the Kolmogorov backward equation
\begin{equation}
\partial_{t^\prime} P(f^\prime, t^\prime) = s \mu f^\prime (1-f^\prime) \partial_{f^\prime} P(f^\prime, t^\prime)  + \mu \frac{ f^\prime (1-f^\prime)}{N} \partial^2_{f^\prime} P(f^\prime, t^\prime),
\label{bkwdsvar}
\end{equation}
where $f^\prime$ is the initial fraction of $A$ organisms at time $t^\prime$.

We can now solve this equation directly for the fixation probability, using the long time limit, with  $P(f^\prime)= \underset{t \to \infty} {\lim} P(1, t | f^\prime, 0)$, and the boundary conditions $P(1)=1$ and $P(0)=0$.

Equation \ref{bkwdsvar} then leads to Kimura's formula for the fixation probability.
\begin{equation}
 P_{fix}(s, N, f)= \frac{1- \exp(- s N f )}{1-\exp(- s N)} .
 \label{kimura}
\end{equation}
Thus, we find that our microscopic rules result in Kimura's formula for the well-mixed case, in agreement with our numerical results. 

To address the question of how Equation \ref{kimura} changes when spatial structure is added, we first turn to the simpler Moran model. In the Moran model, there are a fixed number of organisms, $N$, and at every time step, one organism is selected to reproduce, and another is selected to die, incrementing the fraction of $A$ organisms, $f$, in steps of size $1/N$. To add selective advantage, one type of organism is made more likely to be chosen for reproduction. Upon choosing $A$ for reproduction a fraction $\frac{f(1+s)}{1+sf}$ of the time, and allowing $A$ to die a fraction $f$ of the time, we have
\begin{equation}
C_f^{(1)}= \frac{1}{N} \frac{f(1+s)}{1+sf} (1-f) -\frac{1}{N} \frac{1-f}{1+sf} (f) = \frac{s f (1-f)}{N(1+sf)} \approx \frac{s f (1-f)}{N}, 
\end{equation}
\begin{equation}
C_{ff}^{(2)}=  \frac{1}{N^2}\frac{f (1-f) (2+s)}{1+sf} \approx \frac{ 2 f (1-f)}{N^2}.
\end{equation}
These Kramers-Moyal expansion coefficients are the same as those for our microscopic model (Sections \ref{micro} and \ref{macro}) up to a constant, and give the same fixation probability. Therefore, instead of working with our microscopic model in higher dimensions, we can work with a ring (or lattice) of well-mixed colonies in the Moran model connected by migration.  Because our model and the Moran model have the same coarse-grained evolution equations at the level of approximation we are interested in, we should be able to apply results derived for the spatially structured Moran model to our system (in the limit of large colony size).

Following \citet{maruyama1974}'s treatment of the Moran model, we consider the moment at which the birth/death occurs in a colony $l$ with $N_l$ organisms, a fraction $f_l$ of them of type $A$ ($\sum_l f_l N_l= f N$). We will assume that the total population size is constant, as before, but we do not need to assume that the number of organisms in each cell, or even the number of cells, remains the same in between steps.

The probability that one $A$ organism replaces one $B$ organism in colony $l$ is
\begin{equation}
r^+_l= \frac{f_l(1+s)}{1+sf_l} (1-f_l) .
\end{equation}
The probability that a $B$ organism replaces an $A$ organism is
\begin{equation}
r^-_l= \frac{1-f_l}{1+sf_l} f_l .
\end{equation}
The probability that $f_l$ does not change is
\begin{equation}
r^0_l= 1- r^+_l- r^-_l .
\end{equation}

Upon considering only events that change the number of $A$ by 1, we can write the probability that $A$ increased as the sum over colonies of the product of the probability that an event occurs in a colony with fraction $f_l$ times the probability that that event was a birth,
\begin{equation}
p = \frac{ \sum_l  \frac{N_l}{N} r^+_l} {\sum_l \frac{N_l}{N} r^+_l + \sum_l \frac{N_l}{N} r^-_l}.
\end{equation}
Similarly, the probability that $A$ decreased is given by
\begin{equation}
q = \frac{ \sum_l  \frac{N_l}{N} r^-_l} {\sum_l \frac{N_l}{N} r^+_l + \sum_l \frac{N_l}{N} r^-_l} .
\end{equation}
All terms have the same $l$ dependence, which cancels. We are left with
\begin{align}
p&=\frac{1+s}{2+s},\\
q&=\frac{1}{2+s}.
\end{align}
We have reduced this problem to that of an unfair coin toss, where the global fraction is increased by $1/N$ with probability $p$ and decreased by $1/N$ with probability $q$. The fixation probability is then the classic solution to the gambler's ruin problem, the exit probability of a biased random walker. Surprisingly, $p$ and $q$ are independent of the details of $\{ N_l \}$, meaning that the fixation probability is independent of spatial structure as long as this argument holds. However, this argument does \textit{not} imply any spatial structure independence of the time to fixation. Because we are only considering events that change the global fraction of $A$ organisms, the average time between steps of the random walk varies with $f$. 

To complete Maruyama's argument, we solve for the exit probability of a biased random walker 
\begin{equation}
P(a)= \frac{1-(q/p)^a}{1-(q/p)^N} = \frac{1- \frac{1}{(1+s)^a}}{1- \frac{1}{(1+s)^N}} \approx  \frac{1- \exp(-s a)}{1-\exp(-s N)} ,
\end{equation}
where $a=f^\prime N$ is the initial number of organisms of species $A$. 
This agrees with Equation \ref{kimura}:
\begin{equation}
 P_{fix}(s, N, f)= \frac{1- \exp(- s N f )}{1-\exp(- s N)} .
 \end{equation}

The derivation of the spatial structure independence of Kimura's formula relies on the assumptions that our population can be modeled as a set of well-mixed colonies and that all organisms in the population are equally likely to reproduce/die and cause the global fraction to increase/decrease. These assumptions hold on average for our model when only diffusion is added. Our birth rate is defined to always be equal for all organisms, and the death rate is concentration dependent. Since the diffusive term ensures that the concentration stays more or less uniform, this will give an approximately uniform death rate. Therefore, we expect fixation probabilities in our model to follow Kimura's formula in any dimension when only diffusive motion is included.

These assumptions can fail when a flow field is added. In particular, any flow field that, even temporarily, concentrates organisms will violate the necessary assumptions, as crowded organisms are more likely to die. For a related study of fixation probabilities in a variety of interesting flow fields, see \cite{herrerias}.

\subsection{Random Walk Analysis of Fixation in a Sine Wave Flow}

\begin{figure}[htp]
\includegraphics[width=15cm]{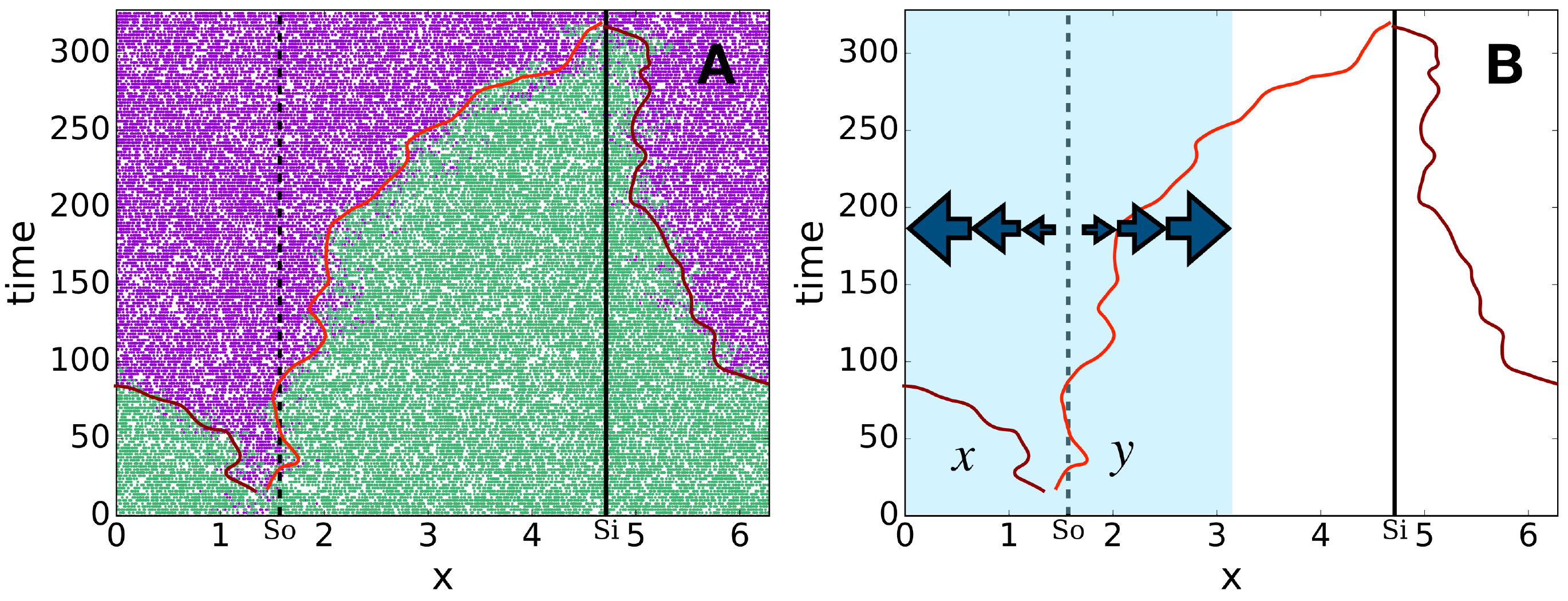}
\caption{{\label{lines} Set-up for random walk approximation in a sine wave flow, with source ($So$) marked by the dashed line and sink ($Si$) marked by the solid line.
(A): One realization of competition between two species in a sine wave flow, with genetic boundaries traced in light/dark red. (B): The half domain used in Figure \ref{rwtri}, with random walkers $x$ and $y$ labeled. Arrows emphasize that the flow magnitude increases away from the source. 
}}
\end{figure}

Suppose we have a sine wave flow field with amplitude $A_0$. Consider the initial condition of Figure 1 in the main text, where a small number of purple organisms start in a narrow window surrounded by green organisms, both species distributed so that the total concentration is the steady state no-flow carrying capacity. Because the concentration is approximately time-independent and spatially uniform in the presence of weakly compressible flows, we can think of fixation events as the convergence of two genetic boundaries, regarded as random walkers on a ring with a spatially varying bias. If the convergence of the P$|$G and G$|$P boundaries is such that the purple sector is pinched shut, this is a fixation event for green, and vice versa (see Figure \ref{lines}.A). Simulations of random walkers confirm that this simple model accurately describes our agent-based simulations. 

With our sine wave flow field, there is a single source, $So$, at $x=\pi/2$ and a single sink, $Si$, at $x=3 \pi /2$. Consider the half domain symmetrically bracketing the source in the periodic domain [0, $2 \pi$], as in Figure \ref{lines}.B. If two random walking genetic domain walls G$|$P and P$|$G exit on opposite sides of this region, it is unlikely that they are able to return. Hence, the probability of exiting on opposite sides of the half domain before converging gives an approximate fixation probability for the purple species. We further approximate the sine wave as a linear velocity field in this half domain. 

This simplified problem can be solved analytically. The two random walkers in the interval $x_s \pm \pi/2$, where $x_s$ is the location of the source, can be mapped to the $x$ and $y$ coordinates of a single random walker in two dimensions, as in Figure \ref{lines}.B. Without loss of generality, we set $x_s=0$ and $y \geq x$, as the walkers are assumed to converge if they cross paths. The exit probabilities for the two-dimensional walker give the approximate fixation probabilities of our system. If the walk reaches the line $x=y$ first, convergence has occurred and green fixes. If the walk reaches the top left corner, where $x=-\pi/2, y=\pi/2$, divergence has occurred and purple fixes. 

The equations for the exit probabilities, $p(x,y)$, are given by a Kolmogorov backward equation that can be derived with a first step analysis. If the random walker begins at the line $x=y$, its chance of first reaching the corner is 0. Therefore, $p(x,x)=0$ gives one boundary condition. Similarly, $p(-\pi/2, \pi/2)=1$. The other parts of the boundary represent the case in which one of the original random walkers has exited the half domain, but the other one remains inside. In this case, we must wait and see where the second walker decides to exit. Therefore, the boundary condition for $p(x, \pi/2)$ and $p(-\pi/2, y)$ is the solution of a one-dimensional diffusion-with-drift problem, with $p(x)$ fixed at 0 and 1 on the appropriate ends. The Kolmogorov backward equation inside the triangle reads
\begin{align}
D \partial_x^2 p(x,y)&+ D \partial_y^2 p(x,y)+ A_0 x \partial_x p(x,y) + A_0 y \partial_y p(x,y)=0,\\
\text{with boundary conditions, \hspace{2 em}}&\\
p(-\pi/2, y)&= \frac{1}{2} + \frac{\text{Erf}\Big[y \sqrt{\frac{A_0}{2D}}\Big]}{2\text{Erf}\Big[\frac{\pi}{2}\sqrt{\frac{A_0}{2D}}\Big]},\\
p(x, \pi/2)&= \frac{1}{2} - \frac{\text{Erf}\Big[x \sqrt{\frac{A_0}{2D}}\Big]}{2\text{Erf}\Big[\frac{\pi}{2}\sqrt{\frac{A_0}{2D}}\Big]},\\
p(x,x)&=0.
\end{align}

This problem has the solution 
\begin{equation}
p(x,y)=\frac{1}{2\text{Erf}\Big[\frac{\pi}{2}\sqrt{\frac{A_0}{2D}}\Big]}\Big(\text{Erf}\Big[y \sqrt{\frac{A_0}{2D}}\Big]-\text{Erf}\Big[x \sqrt{\frac{A_0}{2D}}\Big]\Big).
\label{sol}
\end{equation}

Now we are ready to understand Figure 1 in the main text. Let $y$ be a fixed amount $\Delta$ greater than $x$. Equation \ref{sol} becomes
\begin{equation}
p(x,x+\Delta)=\frac{1}{2 \int_0^{\pi/2} \exp\Big(-\frac{A_0}{2 D} t^2\Big) dt}\Big(\int_x^{x+\Delta}  \exp\Big(-\frac{A_0}{2 D} t^2\Big) dt \Big).
\end{equation}

To first order in $\Delta$, we have
\begin{equation}
p(x,x+\Delta) \approx \frac{\Delta \exp\Big(-\frac{A_0}{2D} x^2\Big)}{2 \int_0^{\pi/2} \exp\Big(-\frac{A_0}{2 D} t^2\Big) dt}.
\end{equation}
This is the Gaussian probability distribution observed in our simulations, with a variance of $D/A_0$. The denominator, upon extending the integration limits to $\pm \infty$, normalizes the numerator, and we are led to Equation 6 in the main text,
\begin{equation}
p(x,x+\Delta) \approx \Delta \cdot \mathcal{N}(x| x_s, D/A_0).
\end{equation}
where $\mathcal{N}$ is the normal distribution. 

\subsection{Turbulent Velocity Field}
We generate a synthetic turbulent velocity field as in Ref. \cite{benzi} using the Sabra shell model. Shells with wavenumbers $k_n=2^{n-1}$, with $n=1,2,...,25$ each have a complex, time-dependent velocity $u_n$. Parameters were chosen to mimic the intermittency of the three dimensional Navier-Stokes equation (free parameter $\delta=0.4$).  

For simulations in which we do not wish to have an identical flow for each realization, a random phase is added to each shell velocity and the model is evolved for twenty times the largest shell turnover time prior to introducing organisms. This protocol ensures that the phases satisfy the equations of motion, and that we obtain a statistically independent velocity field. 

A real space velocity field is obtained through a modified Fourier transform, where we construct the longest wavelength mode out of both $u_1$ and $u_2$ to create a broader palette of flow realizations,
\begin{equation}
u(x,t)= A_0 \Big( \frac{1}{4} \big[u_1 e^{i k_1 x} + u_1^*e^{-i k_1 x}\big] +  \sum_{n=2}^{25} \big[ u_n e^{i k_{n-1} x} + u_n^*e^{-i k_{n-1} x} \big]\Big).
\end{equation}

This procedure produces a Reynolds number of approximately $2 \times 10^6$ for $A_0=1$.

\subsection{The Effect of Turbulent Dynamics}
In our main results (Figures 5 and 6 of the main text), for each independent simulation, we initialize organisms randomly according to a uniform distribution in the correct proportions (10\% purple, 90\% green). In addition to this, each simulation has an independent flow field, which we solve for by adding a random phase as described above. This gives us results that do not depend on the particular initial condition of our shell model. 

However, one way that we can examine the effect of dynamics is to find the fixation probability in the presence of a single flow field time series, rather than an ensemble. In this case, the initial organism positions are still set independently at the beginning of each simulation, but the flow field at each point in time is the same between trials. We can then compare the fixation probabilities for different flows, and try to connect differences in the genetic outcomes to differences in the flows. 

We show an example of this in Figure \ref{icturb}. As in Figure 1 of the main text, we vary the initial location of one particular species and gather fixation statistics as a function of space. The two figures shown are for two specific turbulent time series, each pictured in the inset. We see that when long-lived sources fluctuate about a mean position, successful fixation attempts are localized around these source regions much like in the case of the stationary sine wave, strongly suggesting a reduction in the effective population size. However, when these sources move quickly across the system, successful fixation attempts can originate at many locations, and the effective population size suffers a more modest reduction. 

In general, the average fixation time for the systems we have studied is significantly longer than the largest eddy turnover time in our turbulence model. 

\subsection{Reynolds Number Scaling}
To find the Reynolds number dependence of our effective population size, we focus on the quantity
\begin{equation}
\label{1}
n_s \tau_s \sim \langle (\partial_x v)^2 \rangle^{1/2} \langle  (\partial_x v)^{-2} \rangle^{1/2} . 
\end{equation}
In the main text, neglecting intermittency effects, we argue that we can approximate this quantity as 1. Here, we calculate the effect of intermittency, and find that it introduces a non-trivial scaling with $Re$.

Using the multifractal approach, we assume that the statistical properties of velocity fluctuations at scales $r$ and $R$ ($r < R$) are given by
\begin{equation}
\label{multiscale}
\delta v(r) = \delta v(R) \left[ \frac{r}{R} \right]^h,
\end{equation}

with probability $P_h(r/R) \sim (r/R)^{3-D(h)}$.  We use \ref{multiscale} to write the condition for dissipation effects to become relevant as
\begin{equation}
\label{vergassola1}
\frac{ \delta v(r) r}{ \nu}  =  \frac{ \delta v(R) R}{ \nu}  \left[ \frac{r}{R} \right]^{1+h} \sim 1.
\end{equation}
We label the dissipation scale $\eta(h)$. Letting $R=L$, the system size, we find
\begin{equation}
\label{etah}
\eta(h) =  Re^{-\frac{1}{1+h}} \, L,
\end{equation}
which is true with probability $(\eta(h)/L)^{3-D(h)} = Re^ {-\frac{3-D(h)}{1+h}}$.  

Now, we can use the above formulation to compute the scaling behavior in $Re$ of moments of velocity gradients:
\begin{equation}
\label{gradienti}
\langle (\nabla v)^p \rangle = \int dh \frac{\delta v(h)^p}{\eta(h)^p} Re^{-\frac{3-D(h)}{1+h}}  \sim \left[\frac{\delta v(L)}{L} \right]^p \int dh Re^{ -\frac{p(h-1)+3-D(h)}{1+h}} \sim \left[\frac{\delta v(L)}{L} \right]^p Re^{\chi(p)}
\end{equation}
where 
$$\chi(p) = sup_h [ -(p(h-1)+3-D(h))/(1+h)]$$ 
From eq. (\ref{gradienti}) it follows that
\begin{equation}
n_s \tau_s  \sim Re^{\alpha}
\end{equation}
where
$$
\alpha = \frac{\chi(2)+\chi(-2)}{2}
$$

We can evaluate $\chi(p)$ using our knowledge of the anomalous scaling in homogeneous and isotropic turbulence for small $p$. The final result reads:
\begin{equation}
\label{estimate}
n_s \tau_s  \sim Re^{0.08}
\end{equation}
Eq. (\ref{estimate}) tells us that, beyond the mean field approach in which $n_s \tau_s = \text{const.}$,   a Reynolds number dependence shows up. However, this dependence is very weak. 

A detailed discussion of the multifractal formulation can be found in a recent review paper \cite{benzibiferale}.


\begin{figure}[htp]
\includegraphics[width=6.7cm]{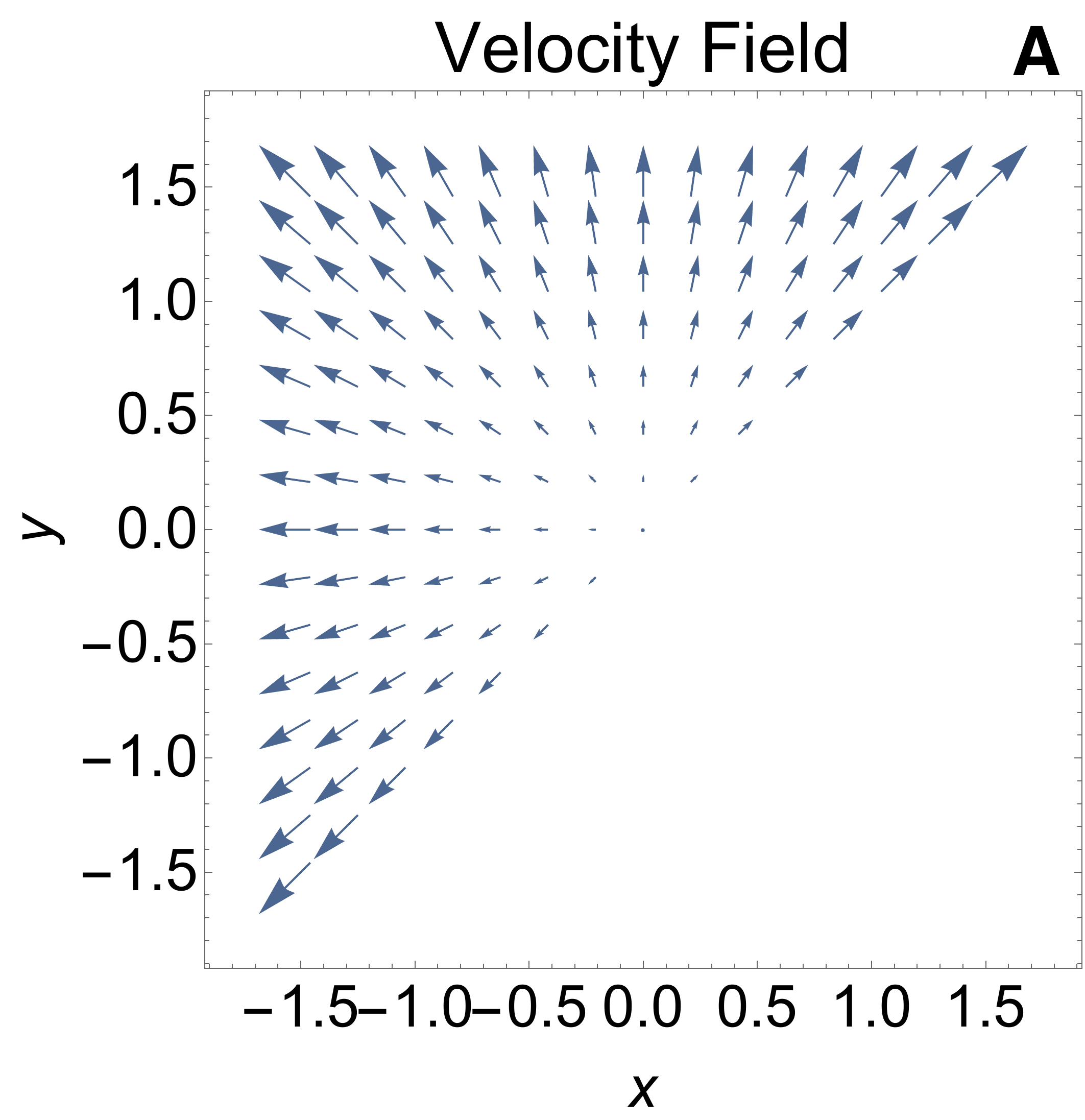}
\includegraphics[width=8.0cm]{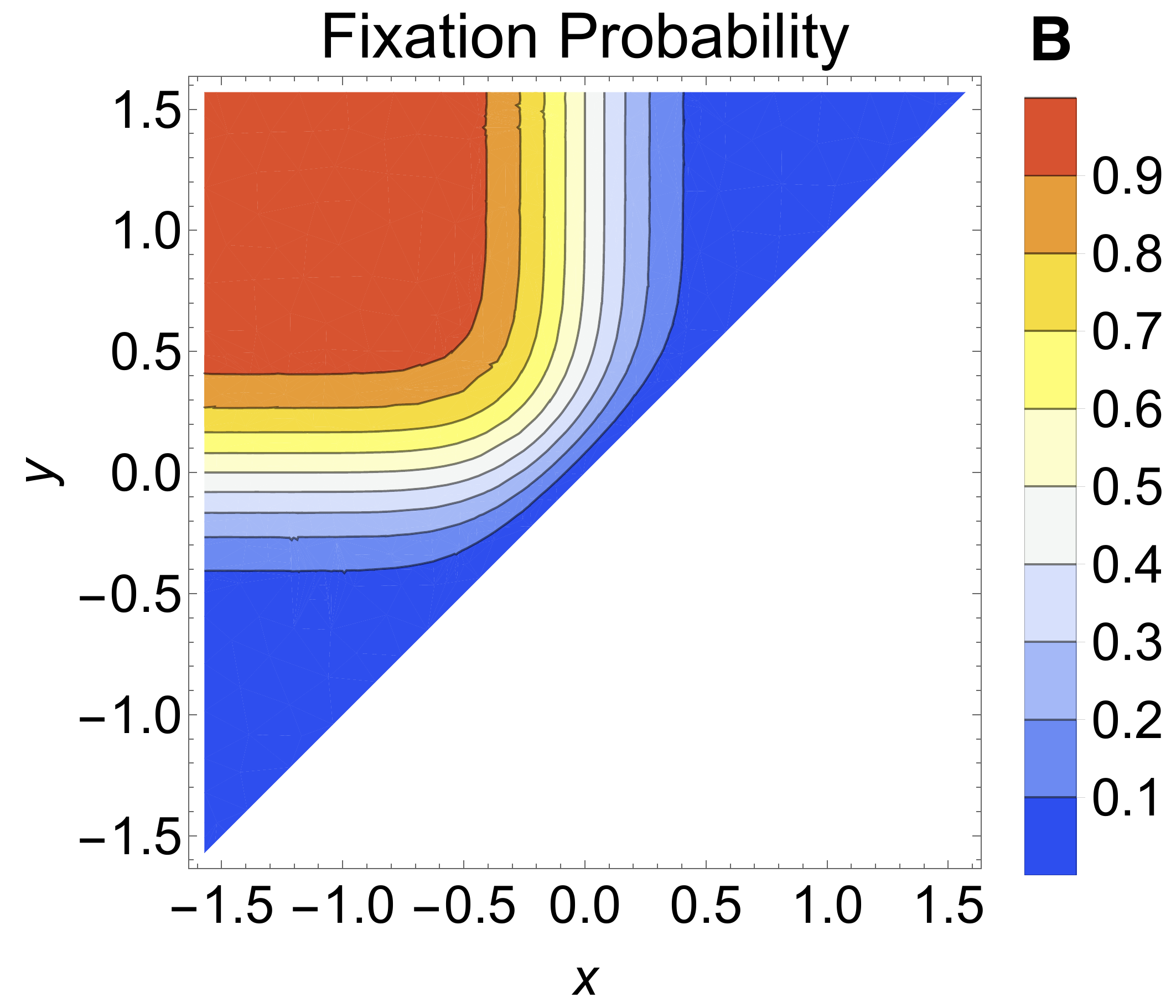}
\caption{{\label{rwtri} (A): Spatially dependent biasing velocity field for a two dimensional random walk problem on a triangular domain. $x \in [-\pi/2, \pi/2]$ and $y \in [x, \pi/2]$. (B): Fixation probability as a function of initial position. 
}}
\end{figure}

\begin{figure}[htp]
\begin{center}
\includegraphics[scale=0.64]{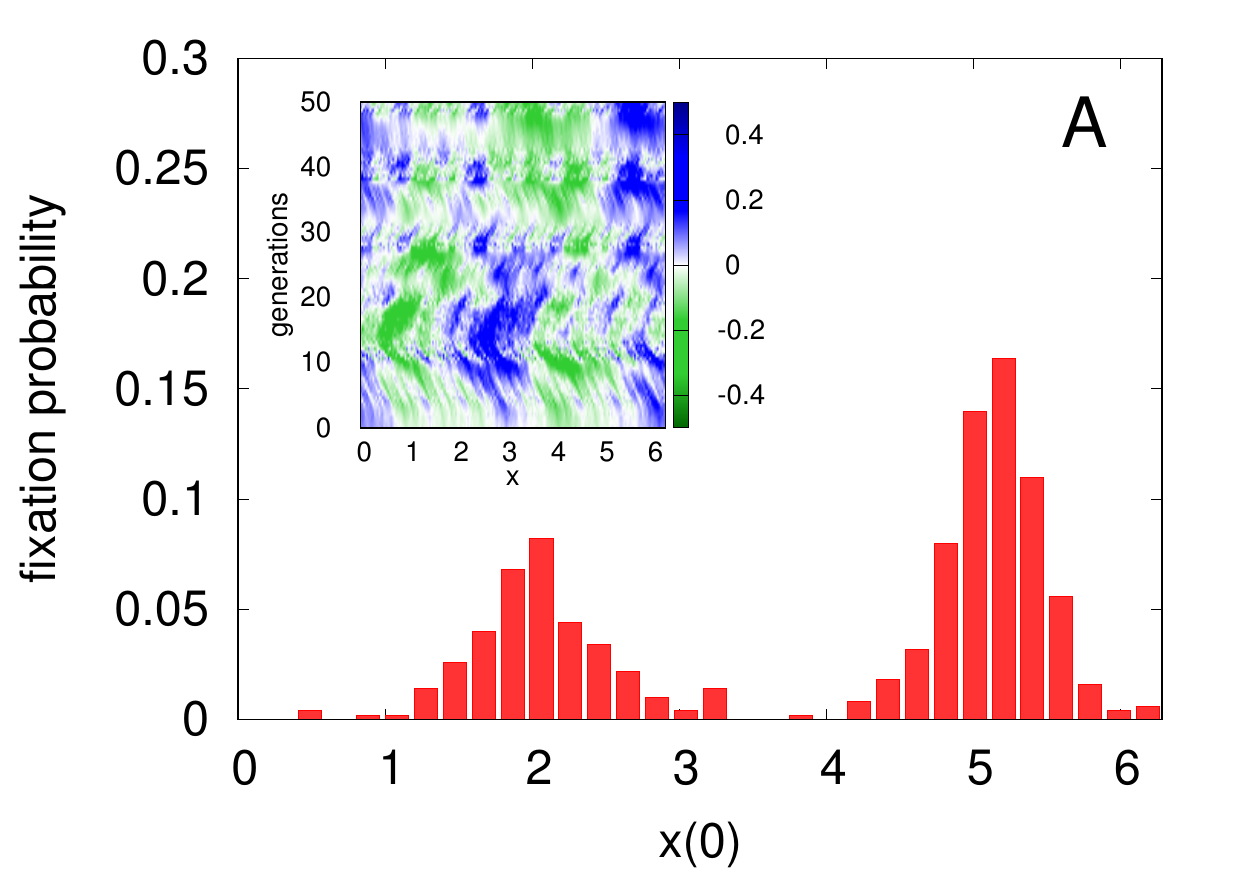}
\includegraphics[scale=0.64]{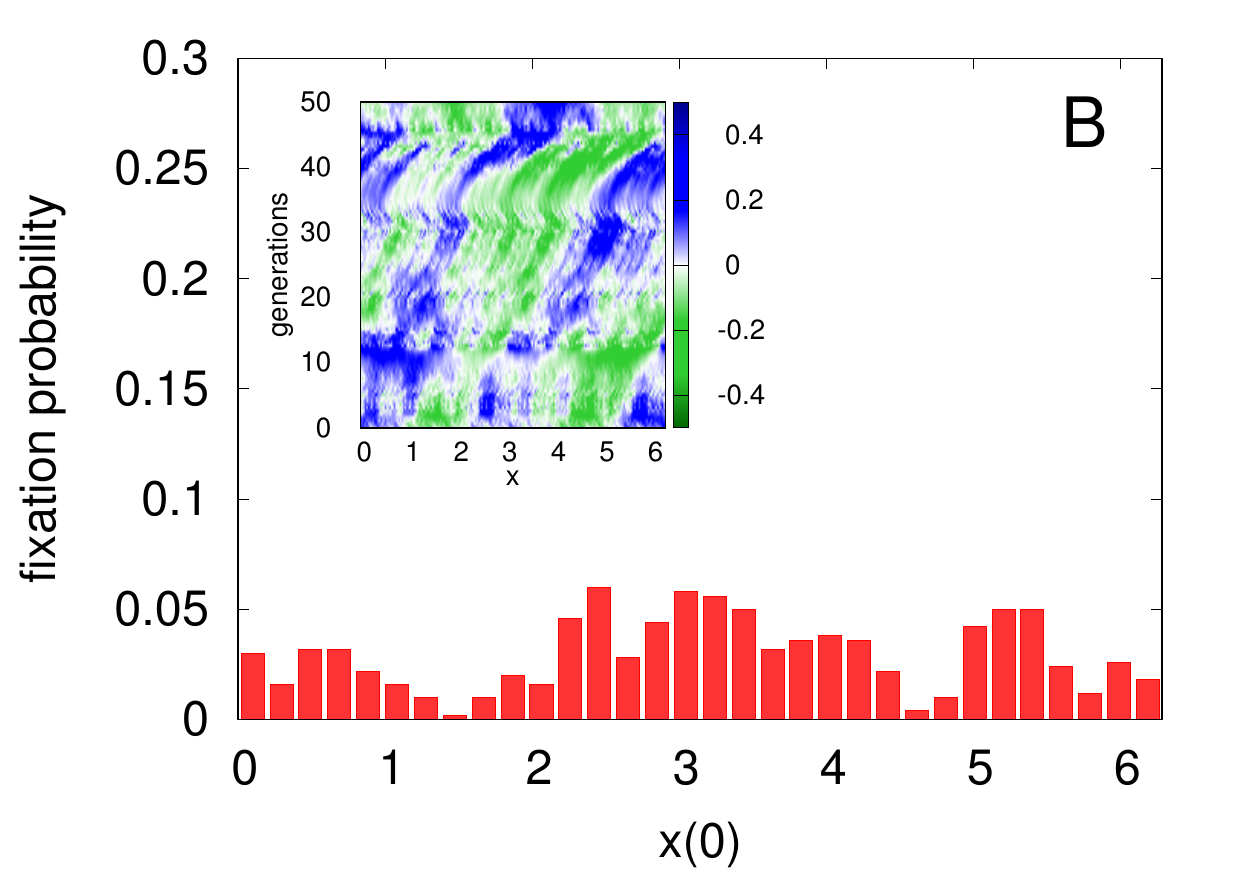}
\caption{{\label{icturb} Fixation probability as a function of initial position for two specific velocity time series generated using the one-dimensional shell model of compressible turbulence. Each figure shows 500 realizations for two neutral species subjected to the turbulent velocity field pictured in the inset. The color bar shows the magnitude of the velocity at each point in time and space. Sources are represented by locations where there is green on the left and blue on the right (G$|$B = a positive slope zero crossing). The locations that sources reach have an enhanced probability of being the origin of a successful fixation attempt. Conversely, fixation is less successful in the vicinity of sinks (= B$|$G) (A): In this case, significant sources fluctuate about a mean position, creating localized source regions. (B): Significant sources in this flow realization move across the system, spreading out the distribution of successful fixation attempts. 
}}
\end{center}
\end{figure}

%
